\documentclass[a4paper, 12pt]{revtex4}
\usepackage[dvips]{graphicx}
\usepackage[]{amsfonts}
\usepackage[]{amssymb}
\usepackage[]{bm}

\begin{document}

	\title{Autonomous Learning by Simple Dynamical Systems with Delayed Feedbacks}

	\author{Pablo Kaluza}
	\affiliation{National Scientific and Technical Research Council \& Faculty of Exact and Natural Sciences, National University of Cuyo, Padre Contreras 1300, 5500 Mendoza, Argentina} 
	\affiliation{Abteilung Physikalische Chemie, Fritz-Haber-Institut der Max-Planck-Gesellschaft, Faradayweg 4-6, 14195 Berlin, Germany.}

	\author{Alexander S. Mikhailov} 
	\affiliation{Abteilung Physikalische Chemie, Fritz-Haber-Institut der Max-Planck-Gesellschaft, Faradayweg 4-6, 14195 Berlin, Germany.}

	\date{\today}

	\begin{abstract}
	A general scheme for construction of dynamical systems able to learn generation of the desired kinds of dynamics through adjustment of their internal structure is proposed. The scheme involves intrinsic time-delayed feedback to steer the dynamics towards the target performance. As an example, a system of coupled phase oscillators, which can by changing the weights of connections between its elements evolve to a dynamical state with the prescribed (low or high) synchronization level, is considered and investigated.
	\end{abstract}

	\maketitle

	Even primitive animals can learn to generate movements and time signals. While this can be easily done by computers too, there is a difference: biological learning does not involve digital encoding and information processing. The brain of an animal can be viewed as a structurally flexible dynamical system able to emulate various other dynamical systems in the outside world \cite{Mikhailov_Synergetics}. Neural universal emulation is achieved through an adaptation process which, on long time scales, is based on the modifications of synaptic connections between neurons.
	
	Whereas the brain is unique and its details are complex, relatively simple systems including networks of coupled phase oscillators \cite{Kuramoto} or model genes \cite{Tsuchiya} are also known to exhibit rich collective dynamics. There are applications, such as the design of electronic circuits or synthetic biology (see, e.g., \cite{Tigges} \cite{Danino}), where systems with prescribed dynamical properties need to be constructed. Can one try to endow simple dynamical systems with the capacity of analog adaptive learning? 

	Generally, the aim of the engineering of dynamical systems is to connect given elements in such a way that the desired performance is realized. With a few elements, such as for small oscillatory genetic circuits, rational construction methods can be employed (see, e.g., \cite{Elowitz}). However, rational design rapidly becomes difficult when the size of a system is increased and a complex combinatorial optimization problem is approached. Combinatorial optimization methods, such as stochastic Metropolis algorithms and simulated annealing, have been used in systems engineering problems \cite{KaluzaPRE, KaluzaEPL, KaluzaCHAOS, KaluzaEPJB, Kobayashi2010, Kobayashi2011, Fujimoto, Inoue, Yanagita2010, Yanagita2012}. Thus, analogs of biological signal transduction networks \cite{KaluzaPRE, KaluzaEPL, KaluzaCHAOS, KaluzaEPJB} and model oscillatory genetic networks with prescribed output patterns or oscillation periods \cite{Kobayashi2010, Kobayashi2011} could be constructed. Moreover, the designed networks could be 
made, through further optimization, robust against local structural perturbations, such as deletion of links or nodes, or against noise \cite{KaluzaPRE, KaluzaCHAOS, KaluzaEPL, KaluzaEPJB}. Model genetic networks, which could generate definite stationary expression patterns and thus imitate processes relevant for biological morphogenesis \cite{Fujimoto} or to produce required temporal responses \cite{Inoue} were developed and investigated. Stochastic Metropolis algorithms with replica exchange were employed to design networks of coupled phase oscillators which exhibit high synchronization despite heterogeneities or noise \cite{Yanagita2010, Yanagita2012}. 

	Stochastic optimization methods consist in running computer evolution of a dynamical model: A structural perturbation, e.g. relocation of an interaction link between two elements, is randomly introduced. The performance of the model after the perturbation is evaluated and compared with its previous behavior. If the performance has been improved, the perturbation is always accepted; it may be also sometimes accepted in the opposite case to avoid intermediate optima. These steps are repeated iteratively until a suitable dynamical system is finally obtained. Because such an optimization procedure is guided by a computer, i.e. by an external agent, it can be classified as representing supervised learning. In contrast to this, an autonomous learning process needs to be realized as a gradual structural adaptation of a system which takes place entirely within its bounds and therefore does not demand external supervision. The difference between supervised and non-supervised learning is extensively discussed in the 
field of artificial intelligence (see, e.g. \cite{AI}).

	The aim of this Rapid Communication is to propose and analyze a simple scheme through which the autonomous learning process can be realized for dynamical systems of various origins. An essential component of the proposed stochastic control scheme is the presence of an internal feedback loop with a time delay. Delayed feedbacks are broadly employed in the control theory \cite{Pyragas, Scholl} to stabilize already existing, but unstable, orbits or to suppress chaos. The function of the feedback is however different in the present study. Essentially, it shall be employed to gradually guide the evolution of a dynamical subsystem in its parameter space along a trajectory which leads to the target performance. 
	
	First, a general formulation of the method is provided and then we demonstrate how it can be applied to a specific problem involving synchronization in an ensemble of coupled phase oscillators.

	We choose a general system with variables $\mathbf{x} = (x_1, x_2, ..., x_N)$ whose dynamics is governed by equations

	\begin{equation}
		\frac{d\mathbf{x}}{dt} = \mathbf{f}(\mathbf{x},\mathbf{w})
		\label{equ_variables_dynamics}
	\end{equation}
	
	\noindent
	which depend on parameters $\mathbf{w} = (w_1, w_2, ..., w_K)$. Suppose that this system is responsible in a selected application for execution of a function which needs to be optimized through evolution of the parameters $\mathbf{w}$. This function can be quantified by a performance vector $\mathbf{R} = (R_1, R_2, ..., R_M)$ and, as we assume, this vector can be expressed as a function (or, generally, a functional) of the system's variables, i.e. $\mathbf{R} = \mathbf{F}(\mathbf{x})$. The ideal target performance of the system (\ref{equ_variables_dynamics}) is known and corresponds to a certain vector $\mathbf{R_0}$. Thus, the deviation $\epsilon$ of the actual performance from the target performance of the system can be defined as $\epsilon = |\mathbf{R} - \mathbf{R_0}|$ or, explicitly, as $\epsilon = |\mathbf{F}(\mathbf{x}) - \mathbf{R_0}|$.

	Our aim is to introduce slow intrinsic dynamics of the parameters $\mathbf{w}$ which would lead to minimization of the deviation $\epsilon$. To do this, we assume that $\mathbf{w}$ are themselves (slow) dynamical variables and their temporal evolution is governed by stochastic differential delay equations

	\begin{equation}
	      \tau \frac{d\mathbf{w}}{dt} = -\Big( \epsilon(t) -\epsilon(t - \Delta) \Big) \Big( \mathbf{w}(t) - \mathbf{w}(t - \Delta) \Big) + \epsilon S \bm{\xi}(t)
	      \label{equ_parameters_dynamics}
	\end{equation}

	\noindent
	where $\bm{\xi}(t) = (\xi_1(t), \xi_2(t),..., \xi_K(t))$ are independent random white noises with $\langle \xi(t) \rangle = 0$ and $\langle \xi_{\alpha}(t) \xi_{\beta}(t')\rangle = 2 \delta_{\alpha \beta} \delta(t-t')$ . The coeffiecient $\tau$ defines the characteristic time scale of the parameter variables $\bm w$; if the time scale of the subsystem (\ref{equ_variables_dynamics}) is chosen as unity, it should satisfy the condition $\tau >> 1$ . The delay time $\Delta$ can be chosen depending on a particular guided system and the condition $\tau >> \Delta >> 1$ has to be satisfied. Note that the noise term in equation (\ref{equ_parameters_dynamics}) is proportional to the deviation from the target performance, it becomes progressively weaker as the target is approached. The coefficient $S$ controls the noise intensity.

	Taken together, equations (\ref{equ_variables_dynamics}) and (\ref{equ_parameters_dynamics}) define an extended dynamical system with combined variables $\bm x$ and $\bm w$. Our conjecture is that, under an appropriate choice of $\tau$, $\Delta$ and $S$, this autonomous system would evolve along an orbit in the space of variables $\bm w$ which leads to minimization of the deviation $\epsilon$ from the target performance.

	The extended system consists of the guided subsystem (\ref{equ_variables_dynamics}) which is controlled by the slow steering subsystem (\ref{equ_parameters_dynamics}). The steering subsystem is persistently sensing the current and the previous performances of the guided subsystem and its deviations $\epsilon$ from the ideal target performance. The evolution equation (\ref{equ_parameters_dynamics}) of the steering subsystem includes noises which may be needed to avoid situations where the evolution trajectory ends in an intermediate minimum.

	To illustrate operation of the proposed method, we apply it to a particular optimization problem: How to connect phase oscillators with different given natural frequencies into a network so that a required level of synchronization is reached? Previously, similar problems were treated by using a stochastic Monte Carlo algorithm with replica exchange \cite{Yanagita2010} (and design of networks of identical phase oscillators with maximal synchronization in presence of noise has also been considered \cite{Yanagita2012}). Now, we want to show how to construct an autonomous dynamical system which would include as its subsystem a network of phase oscillators and independently evolve to a state with a desired degree of synchronization.

	We consider the classical Kuramoto model \cite{Kuramoto} of coupled phase oscillators described by equations

	\begin{equation}
		\frac{d \phi_i}{dt} = \omega_i + \frac{1}{N} \sum_{j=1}^{N} w_{ij} \sin(\phi_j - \phi_i)
		\label{equ_kuramoto}
	\end{equation}

	\noindent 
	where $\phi_i$ is the phase and $\omega_i$ is the natural frequency of oscillator $i$. Additionally, interactions are characterized by weights $w_{ij}$  which may be different for each oscillator pair $(i,j)$. As we assume, $w_{ij} = w_{ji}$ and thus the interactions are symmetric. The weights can be positive or negative. 

	Synchronization is quantified by the Kuramoto order parameter $r(t) = \frac{1}{N} | \sum_{j=1}^{N} \exp(i\phi_j) |$. Because it can fluctuate with time, the degree of synchronization at time $t$ is better characterized by an average over a previous time interval $T$, $R(t) = \frac{1}{T} \int_{t-T}^{t}r(t')dt'$. Note that $R$ can vary from zero to one and the value $R = 1$ corresponds to the state with complete phase synchronization, when the phases of all oscillators are identical.

	Our aim is to construct a dynamical system (\ref{equ_variables_dynamics}-\ref{equ_parameters_dynamics}) which, through its autonomous evolution, would converge to a state with some required synchronization parameter $P$. The variables $x_i$ are now replaced by phases $\phi_i$ and the general equation (\ref{equ_variables_dynamics}) is replaced by (\ref{equ_kuramoto}). The ideal performance is $P$ and the deviation from such target performance is $\epsilon = |P - R(t)|$.

	The dynamics for variables $w_{ij}$ are
	\begin{eqnarray}
		\frac{dw_{ij}(t)}{dt} &=& -K_{\tau} \Big(\epsilon(t) - \epsilon(t-\Delta) \Big) \Big( w_{ij}(t-T) \nonumber \\ 
				      & & - w_{ij}(t-T-\Delta) \Big)  + \lambda\frac{w_{ij}(t)}{v(t)} \Bigg( W - v(t) \Bigg) \nonumber \\
				      & & + S \epsilon(t) \xi_{ij}(t),
		\label{equ_ejemplo_dynamics}
	\end{eqnarray}
	where $\langle \xi_{ij}(t) \rangle = 0$ and $\langle \xi_{ij}(t) \xi_{kl}(t') \rangle = 2 \delta_{ik} \delta_{jl} \delta(t-t')$. We have added to the right side of the equation an additional term which, at sufficiently large values of the coefficient $\lambda$, would ensure that the condition $v \approx W$ is approximately satisfied. The mean absolute weight $v$ is defined as $v(t) = \frac{1}{N(N-1)} \sum_{i,j=1}^{N} | w_{ij}(t) |$. Equations (\ref{equ_kuramoto}-\ref{equ_ejemplo_dynamics}) constitute a closed dynamical system which consists of the guided subsystem (\ref{equ_kuramoto}) and the slowly evolving steering subsystem (\ref{equ_ejemplo_dynamics}). 

	Numerical investigations were performed for a system of $N=10$ coupled oscillators. Natural frequencies $\omega_i$ of the oscillators were uniformly distributed between $-0.3$ and $0.3$, so that we had $\omega_i = (i-1)/15 - 0.3$ for $i = 1,2,..., 10$. If uniform coupling with constant weights $w_{ij}=0.3$ is chosen, simulations of the system (\ref{equ_kuramoto}) with such parameters show that is exhibits phase synchronization with $R(t) \approx 0.3$. Can the extended system (\ref{equ_kuramoto}-\ref{equ_ejemplo_dynamics}) learn to improve synchronization by $100\%$ while keeping the same average absolute weight of connections between the oscillators? 

	To answer this question, we set the target synchronization parameter as $P=0.6$ and the target mean absolute connection weight as $W=0.3$. The other parameters were fixed as $\Delta = 250$, $T=200$, $K_{\tau}=0.001$, $\lambda=0.0011$ and $S=0.005$. For integration of dynamical equations, the explicite Euler method with the time step  $dt=0.01$ was used. The initial condition for the weights was $w_{ij}=W$ and the initial phase for the oscillators were taken at random with an uniform distribution between zero and $2\pi$.

	\begin{figure}[!ht] 
		\begin{center}
			\includegraphics[width=1.0\columnwidth,clip]{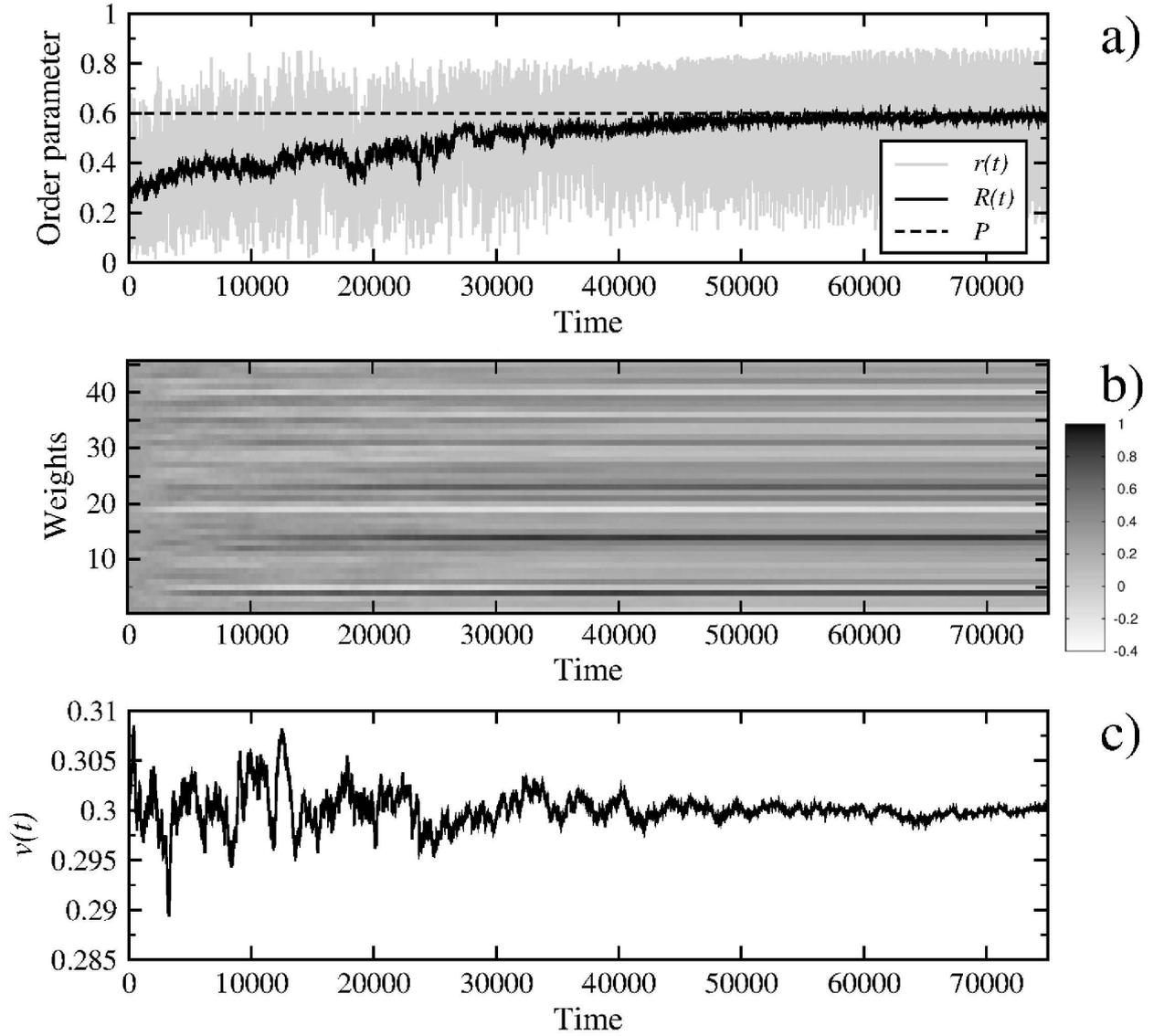}
			\caption{Example of an evolution increasing synchronization. a) Order parameters $r(t)$ (gray) and $R(t)$ (black) as functions of time. b) Weights $w_{ij}$ as functions of time. c) Mean absolute weight $v(t)$ as a function of time.} 
			\label{fig_fig01} 
		\end{center}
	\end{figure}

	Figure \ref{fig_fig01}a shows the computed evolution of the Kuramoto order parameter $r(t)$ and of its average $R(t)$. It can be seen that the mean degree of synchronization, characterized by $R(t)$, increases during the evolution from about $0.3$ to almost $0.6$, as required by the set target value of $P = 0.6$. The instantaneous order parameter $r(t)$ exhibits strong fluctuations around the average value, as should be indeed expected for a system with a relatively small number of oscillators.

	Synchronization improvement is achieved through the redistribution of connection weights which represent slow dynamical variables of the model. Figure \ref{fig_fig01}b shows temporal evolution of all connection weights in the simulation corresponding to Fig. \ref{fig_fig01}a. Because connections are assumed to be symmetric ($w_{ij} = w_{ji}$) and the diagonal elements $w_{ii}$ are zero, there are $10 \times 9/2 = 45$ independent weight variables. They can be enumerated by a single index $\alpha$, i.e. as $w_\alpha$, if we choose $\alpha(i,j) = 45 - (10-i)(10-i+1)/2  + j -i$ with $j>i$. Each horizontal stripe in Fig. \ref{fig_fig01}b displays in gray scale evolution of one of the connection weights. The white color corresponds to the strongest negative weights, whereas the strongest positive weights are shown in black. During the evolution, the average absolute connection weight remains approximately constant and equal to its initial value $v(0) = 0.3$ (Fig. \ref{fig_fig01}c).

	Thus, the system can indeed autonomously steer its evolution to a state with a predefined increased synchronization level. Its dynamics converges to a state where connection weights remain approximately constant and the order parameter fluctuates around the required average value. 

	\begin{figure}[!ht] 
		\begin{center}
			\includegraphics[width=1.0\columnwidth,clip]{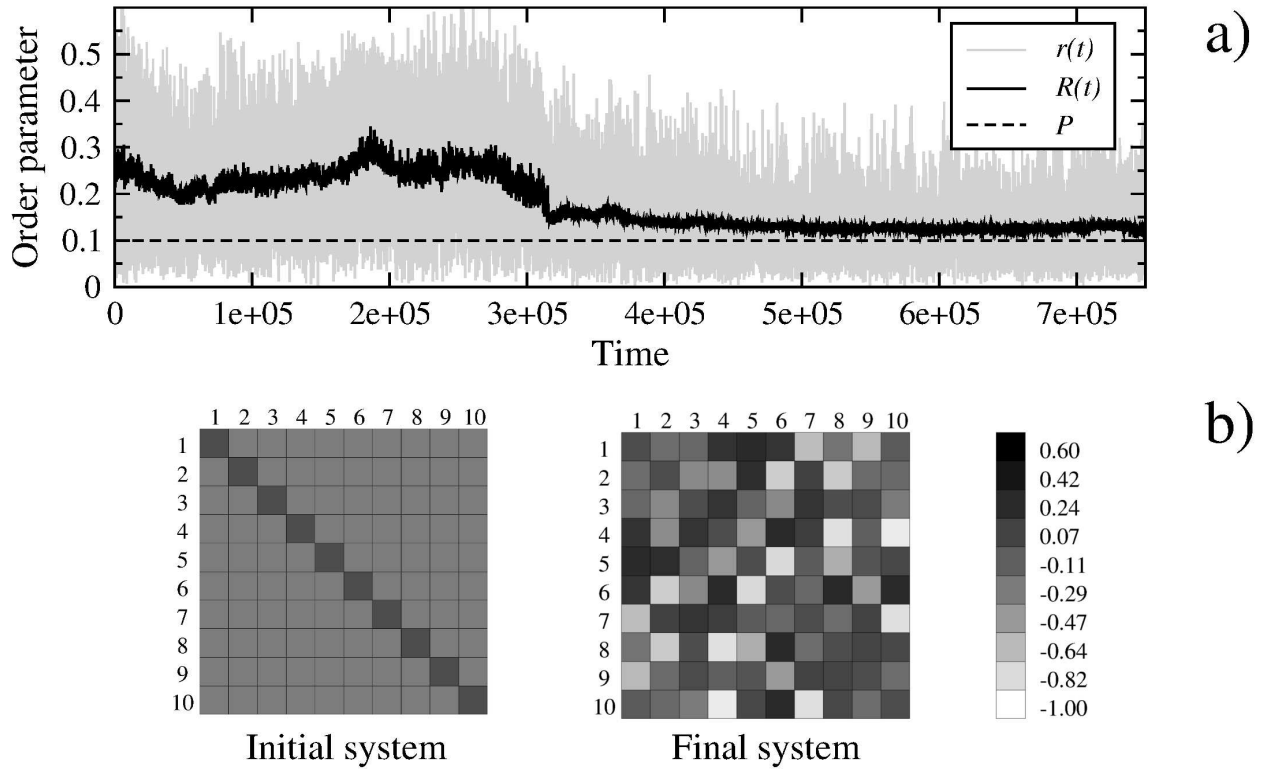}
			\caption{Example of an evolution decreasing synchronization. a) Order parameters $r(t)$ (gray) and $R(t)$ (black) as functions of time. b) Initial and final matrices of connection weights.} 
			\label{fig_fig02} 
		\end{center}
	\end{figure}
	
	In our second example, we show that, by changing the settings, the same network can learn to desynchronize its dynamics. By running simulations of system (\ref{equ_kuramoto}) with equal negative connection weights $w_{ij} = - 0.3$, we find that then its synchronization level is $R \approx 0.27$. Since now our aim is to reduce synchronization, we set $P = 0.1$ as the target level. Starting the simulation with uniform weights $w_{ij} = -0.3$ and keeping the mean absolute weight approximately constant, we obtain an evolution displayed in Fig. \ref{fig_fig02}a. As seen in Fig. \ref{fig_fig02}a, the system is indeed able to reduce its synchronization level from $0.27$ to about $R = 0.1$. This is achieved through the modification of connection weights which acquire both negative and positive values. The initial and final patterns of connection weights are displayed in Fig. \ref{fig_fig02}c.

	\begin{figure}[!ht] 
		\begin{center}
			\includegraphics[width=1.0\columnwidth,clip]{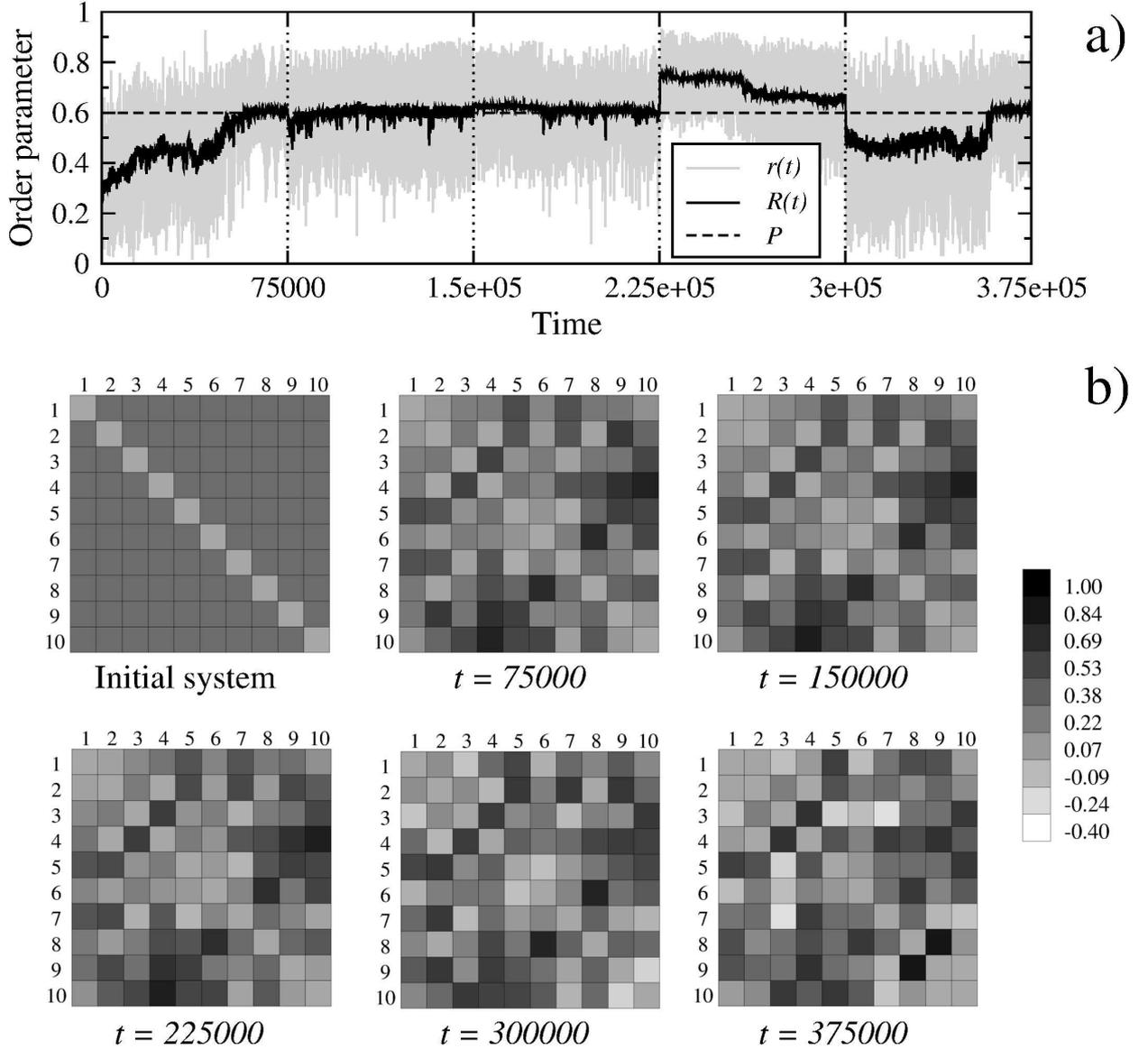}
			\caption{Adaptive learning. The system tends to maintain the required synchronization level $P = 0.6$ despite variations in the natural frequency of the first oscillator (see the text). Evolution of the order parameters $r(t)$ and $R(t)$ (a) and the connection matrices at several time moments (b) are shown.} 
			\label{fig_fig03} 
		\end{center}
	\end{figure}

	The proposed system can not only learn, but also adapt its behavior to the variations of external parameters. Suppose that the frequency $\omega_1$ of the first oscillator changes stepwise with time, taking values $\omega_1 = - 0.3$ for $0 < t < 75000$, $\omega_1 = -0.159$ for $75000 < t < 150000$, $\omega_1 = - 0.282$ for $150000 < t < 225000$, $\omega_1 = 0.145$ for $225 < t < 300000$ and $\omega_1 = 0.067$ for $300000 < t < 375000$. As seen in Fig. \ref{fig_fig03}, the system is able to adapt to such variations, recovering after each change the required synchronization level of $P = 0.6$. It does this by rearrangement of connection weights between the oscillators (Fig. \ref{fig_fig03}b).

	The above examples of coupled oscillators provide a demonstration of how the formulated general learning scheme can operate for a specific model. We have also tested the scheme by applying it to several other models, including genetic expression networks. These further applications, as well as a more detailed analysis for the system of coupled phase oscillators, will be reported in a separate publication.
	
	The autonomous learning scheme, described by equations (\ref{equ_variables_dynamics}) and (\ref{equ_parameters_dynamics}), is employing delayed internal feedbacks. Note however that the use of delayed interactions can be avoided. With this purpose, an additional set of slow dynamical variables can be introduced, so that the dynamics of system parameters (such as, e.g., the connection weights) becomes inertial. The respective study will also be separately reported.
	
	So far, the developed learning scheme has been checked only in numerical simulations. Because of its simplicity, it can be however easily implemented at the hardware level. An interesting related question is whether similar simple schemes of autonomous learning are actually already employed by some biological systems. Biological organisms and individual living cells include a great variety of dynamical systems. Can autonomous learning perhaps take place at the primary level of such biological dynamical systems, not involving at all the neural information processing? It can be particuarly intriguing to check whether regulatory genetic networks can possess this property.
	
	Financial support from the Volkswagen Foundation and from DFG (Germany) through SFB 910 is acknowledged.


\begin{thebibliography}{99}
 
	
	\bibitem{Mikhailov_Synergetics} A. S. Mikhailov, \textit{Foundations of Synergetics I. Distributed Active Systems}, (Springer, Berlin, 1990). 
	
	\bibitem{Kuramoto}  Y. Kuramoto, \textit{Chemical Oscillations, Waves and Turbulence}, (Springer, New York, 1984).
	
	\bibitem{Tsuchiya}  M. Tsuchiya \textit{et al.}, PLoS ONE \textbf{9}, e97411 (2014).
	
	\bibitem{Tigges} M. Tigges \textit{et al.}, Nature \textbf{457}, 516 (2008). 
	
	\bibitem{Danino} T. Danino \textit{et al.}, Nature \textbf{463}, 326 (2009).

	\bibitem{Elowitz} M. B. Elowitz. and S. Leibler, Nature \textbf{403}, 335 (2000).

	\bibitem{KaluzaPRE} P. Kaluza, M. Ipsen, M. Vingron, and A. S. Mikhailov. Phys. Rev. E. \textbf{75}, 015101 (2007).

	\bibitem{KaluzaEPL} P. Kaluza, and A. S. Mikhailov. Europhys. Lett. \textbf{79}, 48001 (2007).

	\bibitem{KaluzaCHAOS} P. Kaluza, M. Vingron, and A. S. Mikhailov. Chaos \textbf{18}, 026113 (2008).

	\bibitem{KaluzaEPJB} P. Kaluza, and A.S. Mikhailov. Eur. Phys. J. B \textbf{85}, 129 (2012).

	\bibitem{Kobayashi2010} Y. Kobayashi, T. Shibata, Y. Kuramoto, and A. S. Mikhailov. Eur. Phys. J. B \textbf{76}, 167-178 (2010).
	
	\bibitem{Kobayashi2011} Y. Kobayashi, T. Shibata, Y. Kuramoto, A. S. Mikhailov. Phys. Rev. E \textbf{83}, 060901 (2011).
      
	\bibitem{Fujimoto} K. Fujimoto, S. Ishihara, and K. Kaneko, PLoS ONE \textbf{3}, e2772 (2007).
  
	\bibitem{Inoue}  M. Inoue, and K. Kaneko. PLoS Comput Biol \textbf{9}(4), e1003001 (2013).

	\bibitem{Yanagita2010} T. Yanagita, and A. S. Mikhailov. Phys. Rev. E \textbf{81}, 056204 (2010). 

	\bibitem{Yanagita2012} T. Yanagita, and A. S. Mikhailov. Phys. Rev. E \textbf{85}, 056206 (2012).
    
	\bibitem{AI} M. Mohri, A. Rostamizade, and A. Talwalkar, \textit{Foundations of Machine Learning}, (MIT Press, 2012).

	\bibitem{Pyragas}  K. Pyragas, Phys. Lett. A \textbf{170}, 421 (1992).
	
	\bibitem{Scholl} E. Sch\"oll, and H. G. Schuster (eds), \textit{Handbook of Chaos Control: Foundations and Applications}, (Wiley-VCH, Weinheim, 2nd revised edition, 2007).

	
			

\end{thebibliography}
\end{document}